\documentclass[reqno, 12pt]{amsart}
\usepackage[ansinew]{inputenc}
\usepackage{amsfonts}
\usepackage{latexsym}
\usepackage{amsmath}
\usepackage{graphicx}

\newtheorem{defin}{Definition}[section]

\begin{document}

\footnote{*Corresponding author: Jian-Guo Liu, Email:
20101059@jxutcm.edu.cn. }

\begin{center}{\bf \large Multiple rogue wave solutions for the generalized
(2+1)-dimensional Camassa-Holm-Kadomtsev-Petviashvili equation}
\end{center}

\vskip 4mm

\begin{center} Jian-Guo Liu$^{1}$, Huan Zhao$^{2}$
\end{center} \vskip 2mm

\noindent$^{1}$ College of Computer, Jiangxi University of
Traditional Chinese Medicine, Jiangxi 330004, China\\
\noindent$^{2}$ Nanchang Institute of Technology, Jiangxi 330044, China
 \vskip 6mm

{\bf Abstract:}\,Based on the symbolic computation approach,
multiple rogue wave solutions of the generalized (2+1)-dimensional
Camassa-Holm-Kadomtsev-Pet-viashvili equation are studied. As an
example, we present the 1-rogue wave solutions, 3-rogue wave
solutions and 6-rogue wave solutions. Furthermore, some dynamics
features of the obtained multiple rogue wave solutions are shown by
3D, contour and density graphics. \vskip 2mm
 {\bf Keywords:}\,
Rogue wave, dynamics features, Camassa-Holm-Kadomtsev-Pet-viashvili
equation. \vskip 2mm

\vskip 4mm

\noindent {\bf 1. Introduction}\\

\quad Rogue waves (named freak waves or monster waves) are a kind
of ocean waves with strong destructive power. Its peak height is
more than twice that of other waves [1]. Rogue waves can burst out
huge energy in a specific space-time range and then disappear. Rogue
waves are localized in a certain space-time and will not spread [2].
The generation of rogue waves has a lot to do with modulation
instability. In addition to appearing in the ocean, rogue wave can
also be used in optical fibres, Bose-Einstein condensates, super
fluids, atmosphere and so on [3]. As a result, rogue waves have
attracted the attention of many scholars [4-10]. Rogue wave
solutions of many nonlinear evolution equations are obtained with
the help of symbolic computation [11-16].

\quad In this paper, we investigate the following generalized
(2+1)-dimensional Camassa-Holm-Kadomtsev-Petviashvili (gCHKP)
equation [17]
\begin{eqnarray}
(u_t+\alpha u_x+\beta u u_x+\Upsilon u_{xxt})_x+u_{yy}=0,
\end{eqnarray}where $\alpha$, $\beta$ and $\Upsilon$ are arbitrary real
constants. This model represents the role of dispersion in the
formation of patterns in liquid drops. Biswas [18] presented the
1-Soliton solution of Eq. (1) by the solitary wave ansatz. Ebadi
obtained exact solutions based on the $G'/G$ method and exponential
function method [19]. Symmetry reductions were studied by the
classical Lie group method [20]. Low order rogue waves were derived
by Qin [21]. Recently, Osman [17] studied the lump wave  and
multi-soliton solutions of Eq. (1). However, multiple rogue wave
solutions have not been found in other literatures, which will be
our main work in this paper.

\quad The organization of this paper is as follows. Section 2
obtains the 1-rogue wave solutions based on the symbolic computation
approach; Section 3 presents the 3-rogue wave solutions; Section 4
gives the 6-rogue wave solutions; Section 5 makes this conclusions.\\

\noindent {\bf 2. 1-rogue wave solutions}\\

\quad Based on the symbolic computation approach [22-26], suppose
{\begin{eqnarray}\upsilon=x+t, \beta=6 \Upsilon,
u=\tau_0+2\,[ln\xi(\upsilon,y)]_{\upsilon\upsilon},\end{eqnarray}
}Eq. (1) can be reduced to the bilinear form{\begin{eqnarray}
&-&\left(\alpha +6 \tau _0 \Upsilon +1\right) \xi_\upsilon^2+\xi
[\left(\alpha +6 \tau _0 \Upsilon +1\right)
\xi_{\upsilon\upsilon}+\Upsilon
\xi_{\upsilon\upsilon\upsilon\upsilon}+\xi_{yy}]\nonumber\\&+&3
\Upsilon \xi_{\upsilon\upsilon}^2-4 \Upsilon  \xi_\upsilon
\xi_{\upsilon\upsilon\upsilon}-\xi_y^2=0.
\end{eqnarray}
}To obtain the 1-rogue wave solutions, we select {\begin{eqnarray}
\xi(\upsilon,y)=(\upsilon -\mu )^2+\vartheta _1 (y-\nu )^2+\vartheta
_0,
\end{eqnarray}
}where $\mu$, $\nu$, $\vartheta _0$ and $\vartheta _1$ are unknown
real constants. Substituting Eq. (4) into Eq. (3) and equating the
coefficients of all powers $\upsilon$ and $y$ to zero, we have
\begin{eqnarray}
\vartheta_0=-\frac{3 \Upsilon }{\alpha +6 \tau _0 \Upsilon +1},
\vartheta_1=\alpha +6 \tau _0 \Upsilon +1.
\end{eqnarray}Substituting Eq. (4) and Eq. (5) into Eq. (2), the 1-rogue wave
solutions for Eq. (1) can be got as {\begin{eqnarray} u=\tau
_0+\frac{4 [-\frac{3 \Upsilon }{\alpha +6 \tau _0 \Upsilon +1}-(\mu
-\upsilon )^2+(\alpha
   +1) (y-\nu )^2+6 \tau _0 \Upsilon  (y-\nu )^2]}{[-\frac{3 \Upsilon }{\alpha +6 \tau _0
   \Upsilon +1}+(\mu -\upsilon )^2+(y-\nu )^2 \left(\alpha +6 \tau _0 \Upsilon +1\right)]{}^2},
\end{eqnarray}}where $\tau _0$ is an arbitrary constant. Rogue wave
(6) has five critical points as $(\mu\pm\frac{3 \Upsilon
}{\sqrt{-\Upsilon \left(\alpha +6 \tau _0 \Upsilon +1\right)}},
\nu)$, $(\mu, \nu)$ and $(\mu, \nu \pm\frac{\sqrt{3} \Upsilon
}{\sqrt{\Upsilon  \left(\alpha +6 \tau _0 \Upsilon
+1\right){}^2}})$. $(\mu, \nu)$ is the center. When $\Upsilon<0$,
Dynamics features of 1-rogue wave solutions are described in Fig. 1,
which contains 3D, contour and density graphics. When $\Upsilon>0$,
Dynamics features of 1-rogue wave solutions are displayed in Fig. 2.

\includegraphics[scale=0.35,bb=20 270 10 10]{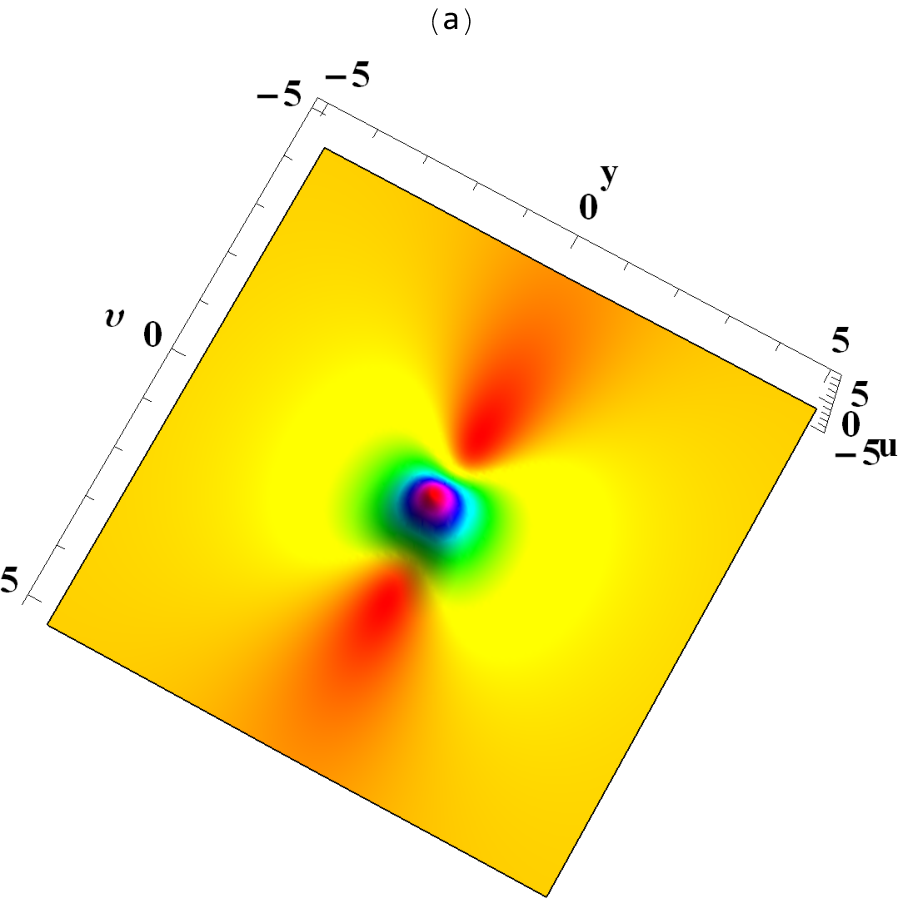}
\includegraphics[scale=0.3,bb=-400 320 10 10]{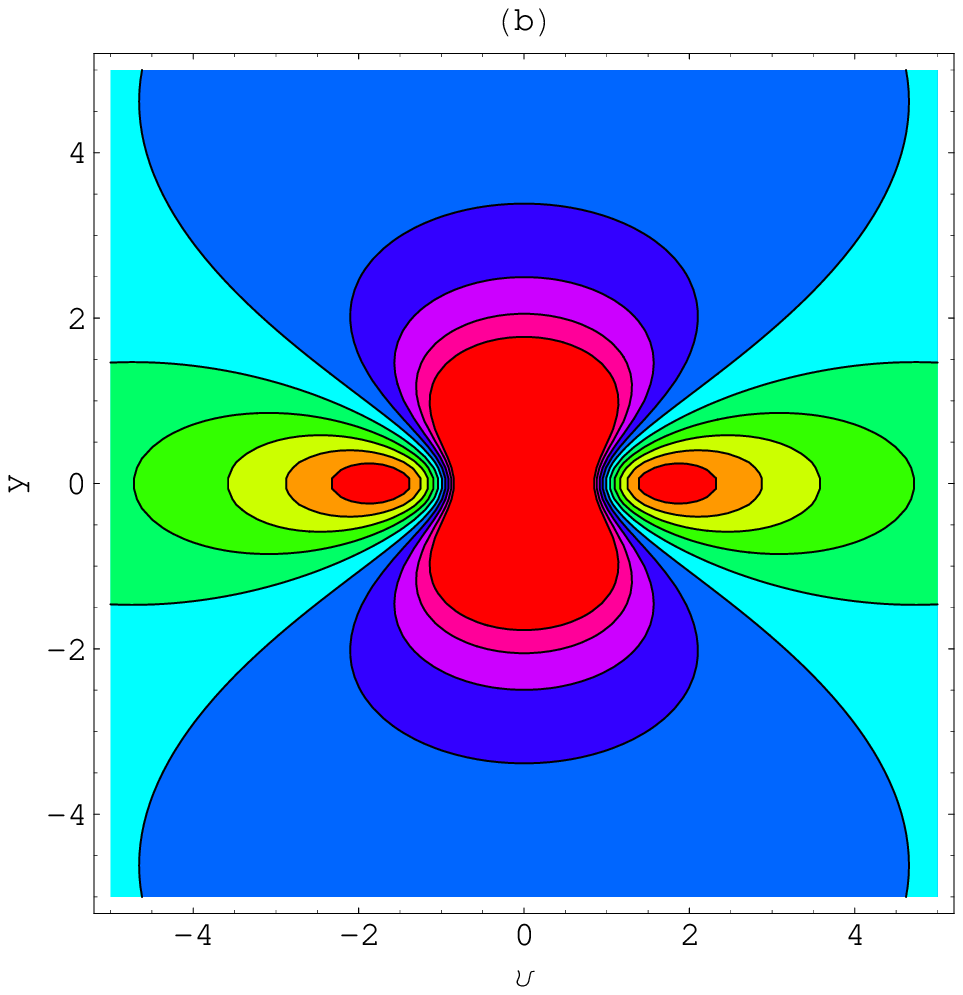}
\includegraphics[scale=0.3,bb=-500 320 10 10]{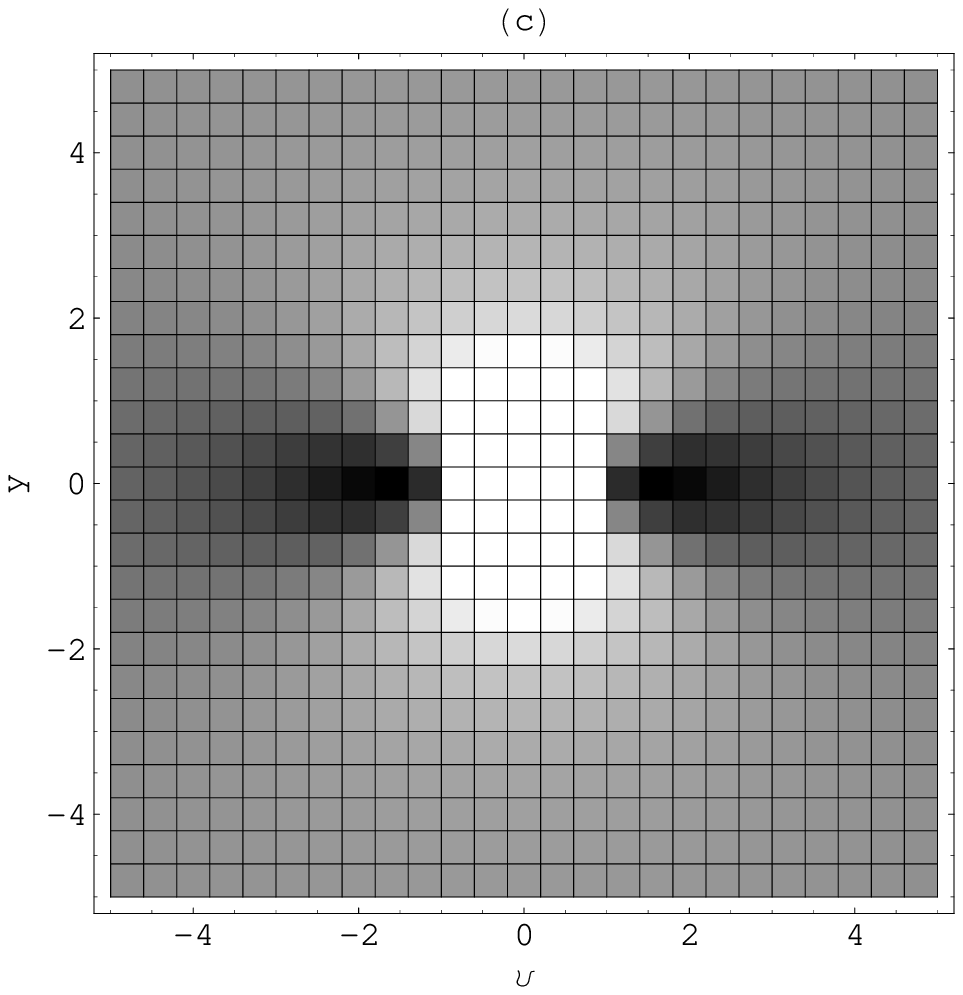}
\vspace{3.2cm}
\begin{tabbing}
\textbf{Fig. 1}. Rogue wave (6) with $\tau_0=\mu=\nu=0$, $\alpha=2$,
$\Upsilon=-1$, (a) 3D graphic,\\  (b) contour plot, (c) density
graphic.
\end{tabbing}

\includegraphics[scale=0.35,bb=20 270 10 10]{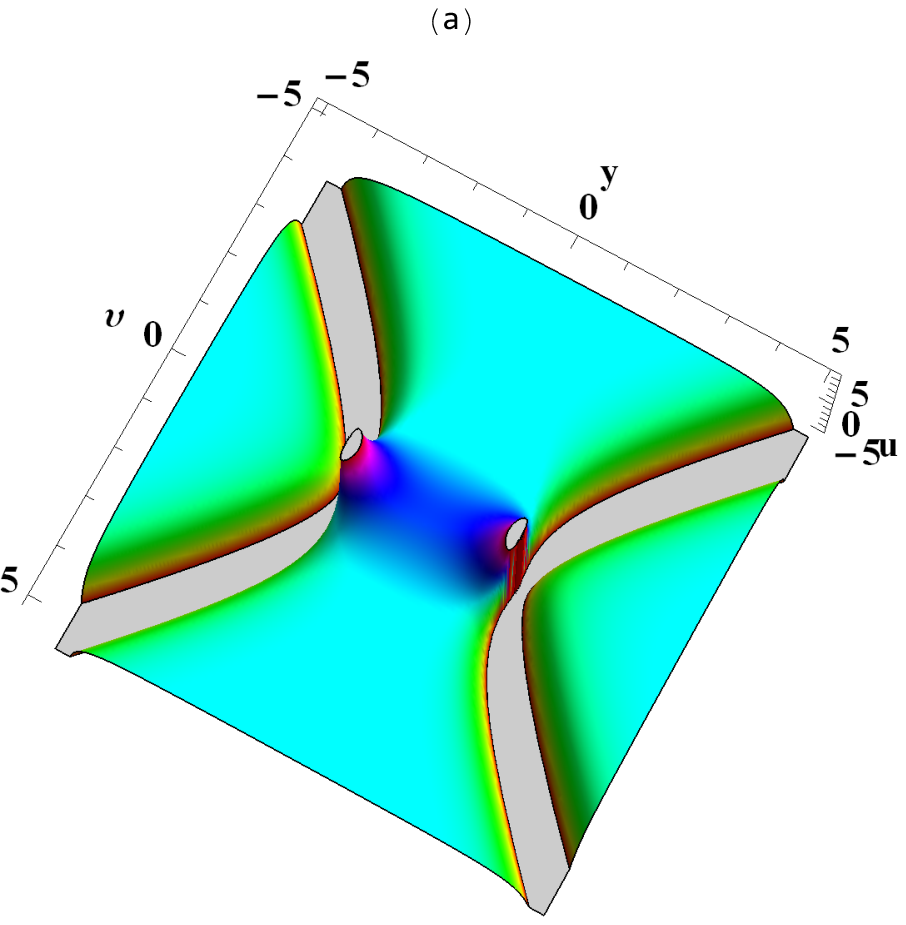}
\includegraphics[scale=0.3,bb=-400 320 10 10]{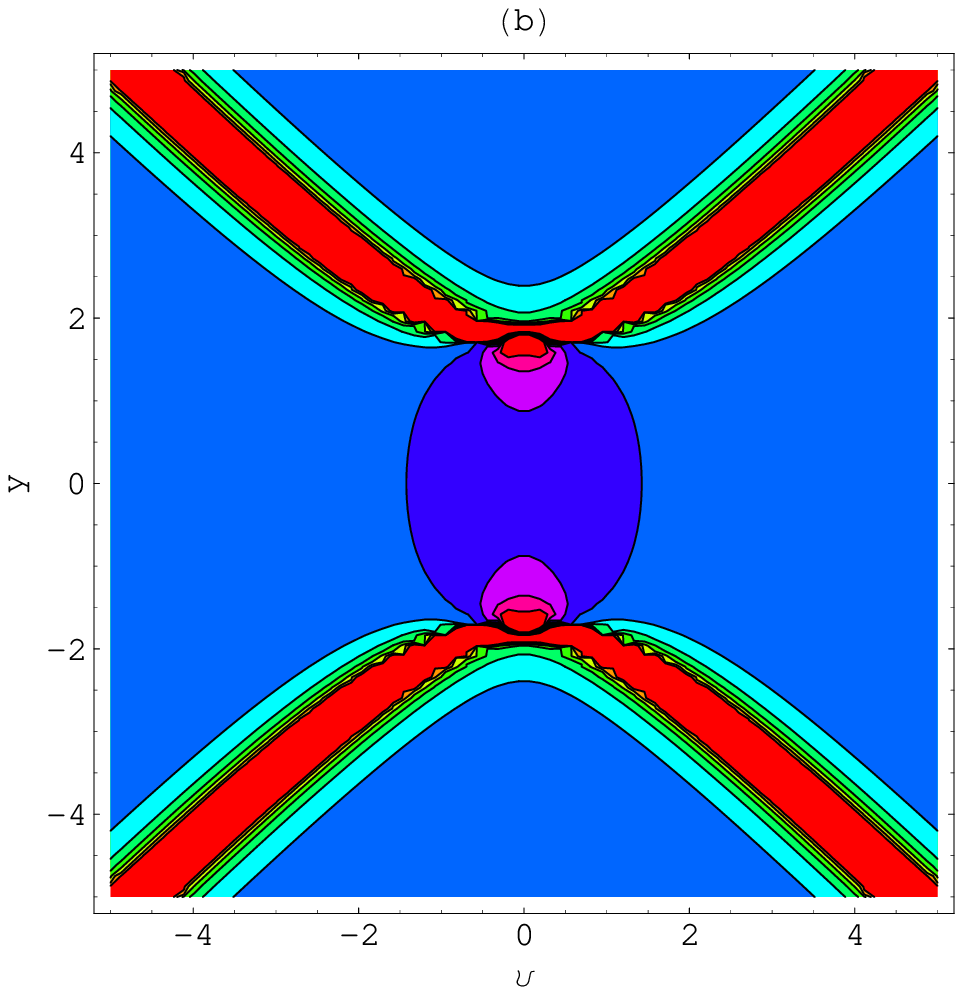}
\includegraphics[scale=0.3,bb=-500 320 10 10]{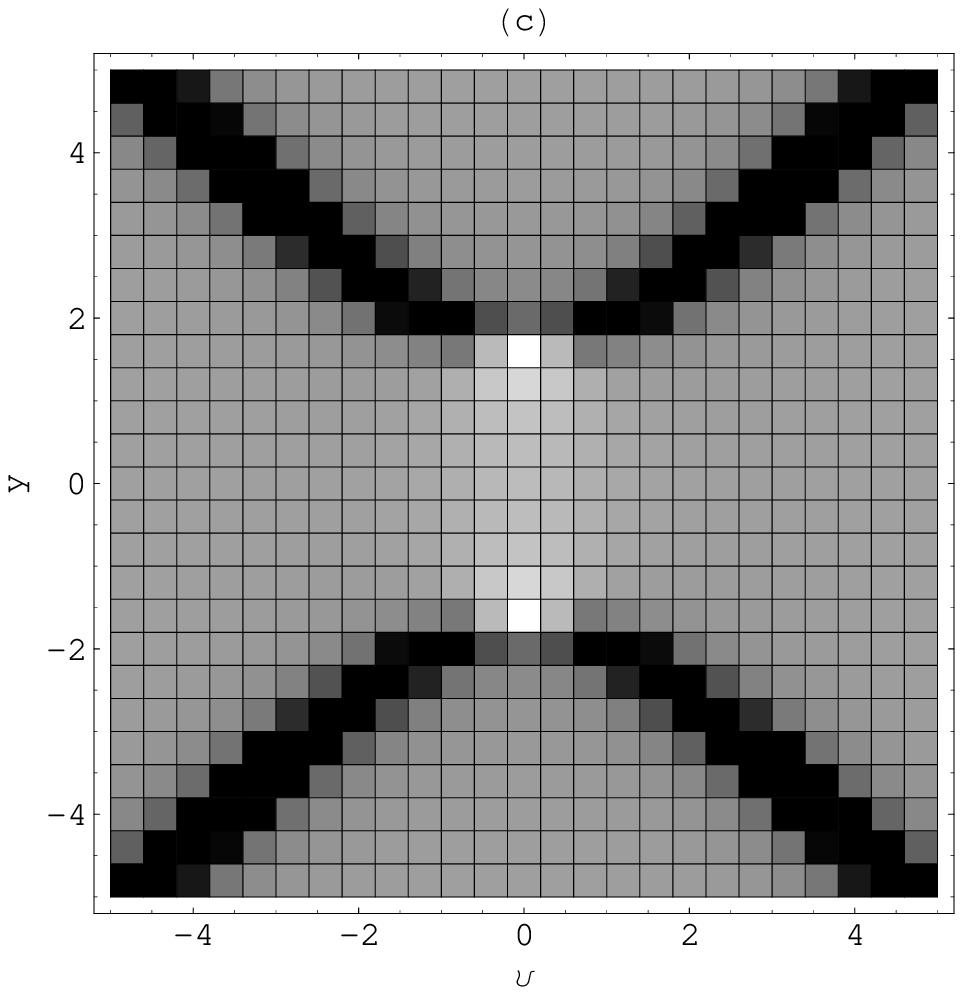}
\vspace{3.2cm}
\begin{tabbing}
\textbf{Fig. 2}. Rogue wave (6) with $\tau_0=\mu=\nu=0$,
$\alpha=-2$, $\Upsilon=1$, (a) 3D graphic,\\  (b) contour plot, (c)
density graphic.
\end{tabbing}

\noindent {\bf 3. 3-rogue wave solutions}\\

\quad To derive the 3-rogue wave solutions, we select
{\begin{eqnarray} \xi(\upsilon,y)&=&\mu ^2+\nu ^2+\upsilon ^6+y^6
\vartheta _{17}+y^4 \vartheta _{16}+2 \mu  \upsilon  \left(y^2
   \vartheta _{23}+\upsilon ^2 \vartheta _{24}+\vartheta _{22}\right)\nonumber\\&+&2 \nu  y \left(y^2 \vartheta
   _{20}+\upsilon ^2 \vartheta _{21}+\vartheta _{19}\right)+\upsilon ^4 y^2 \vartheta _{11}+y^2
   \vartheta _{15}\nonumber\\&+&\upsilon ^2 \left(y^4 \vartheta _{14}+y^2 \vartheta _{13}+\vartheta
   _{12}\right)+\upsilon ^4 \vartheta _{10}+\vartheta _{18},
\end{eqnarray}
}where $\vartheta _i (i=10,\cdots, 24)$  is unknown real constant.
Substituting Eq. (7) into Eq. (3) and equating the coefficients of
all powers $\upsilon$ and $y$ to zero, we get
\begin{eqnarray}
\vartheta_{10}&=&-\frac{25 \Upsilon }{\alpha +6 \tau _0 \Upsilon
+1}, \vartheta_{11}=3 \left(\alpha +6 \tau _0 \Upsilon +1\right),
\vartheta_{14}=3 \left(\alpha +6 \tau _0 \Upsilon
+1\right){}^2,\nonumber\\ \vartheta_{13}&=&-90 \Upsilon,
\vartheta_{23}=-\frac{3 \vartheta _{22} \left(\alpha +6 \tau _0
\Upsilon +1\right){}^2}{\Upsilon }, \vartheta_{16}=-17 \Upsilon
\left(\alpha +6 \tau _0 \Upsilon +1\right),\nonumber\\
\vartheta_{17}&=&\left(\alpha +6 \tau _0 \Upsilon +1\right){}^3,
\vartheta_{20}=\frac{\vartheta _{19} \left(\alpha +6 \tau _0
\Upsilon +1\right){}^2}{5 \Upsilon }, \vartheta_{15}=\frac{475
\Upsilon ^2}{\alpha +6 \tau _0 \Upsilon +1},\nonumber\\
\vartheta_{24}&=&\frac{\vartheta _{22} \left(\alpha +6 \tau _0
\Upsilon +1\right)}{\Upsilon },  \vartheta_{12}=-\frac{125 \Upsilon
^2}{\left(\alpha +6 \tau _0 \Upsilon +1\right){}^2},\nonumber\\
\vartheta_{18}&=&[25 [\mu ^2 \vartheta _{22}^2 \left(\alpha +6 \tau
_0 \Upsilon +1\right){}^5-\Upsilon ^2
   [18 \tau _0 \Upsilon  \left(\mu ^2+\nu ^2\right) [6 \tau _0 \Upsilon  \left(\alpha +2
   \tau _0 \Upsilon +1\right)\nonumber\\&+&(\alpha +1)^2]+(\alpha +1)^3 \left(\mu ^2+\nu ^2\right)+1875
   \Upsilon ^3]]+\nu ^2 \vartheta _{19}^2 \left(\alpha +6 \tau _0 \Upsilon
   +1\right){}^4]\nonumber\\&/&[25 \Upsilon ^2 \left(\alpha +6 \tau _0 \Upsilon
   +1\right){}^3],
\vartheta_{21}=-\frac{3 \vartheta _{19} \left(\alpha +6 \tau _0
\Upsilon +1\right)}{5 \Upsilon }.
\end{eqnarray}Substituting Eq. (7) and Eq. (8) into Eq. (2), the 3-rogue wave
solutions for Eq. (1) can be read as {\begin{eqnarray} u=-\frac{2
\xi_\upsilon^2}{\xi ^2}+\frac{2 \xi_{\upsilon\upsilon}}{\xi
   }+\tau _0,
\end{eqnarray}}where $\xi$ satisfies Eq. (7) and Eq. (8). Dynamics features of
3-rogue wave solutions are shown in Fig. 3 ($\Upsilon<0$) and Fig.4
($\Upsilon>0$). It's very clear that there are three 1-rogue wave
solutions.

\includegraphics[scale=0.35,bb=20 270 10 10]{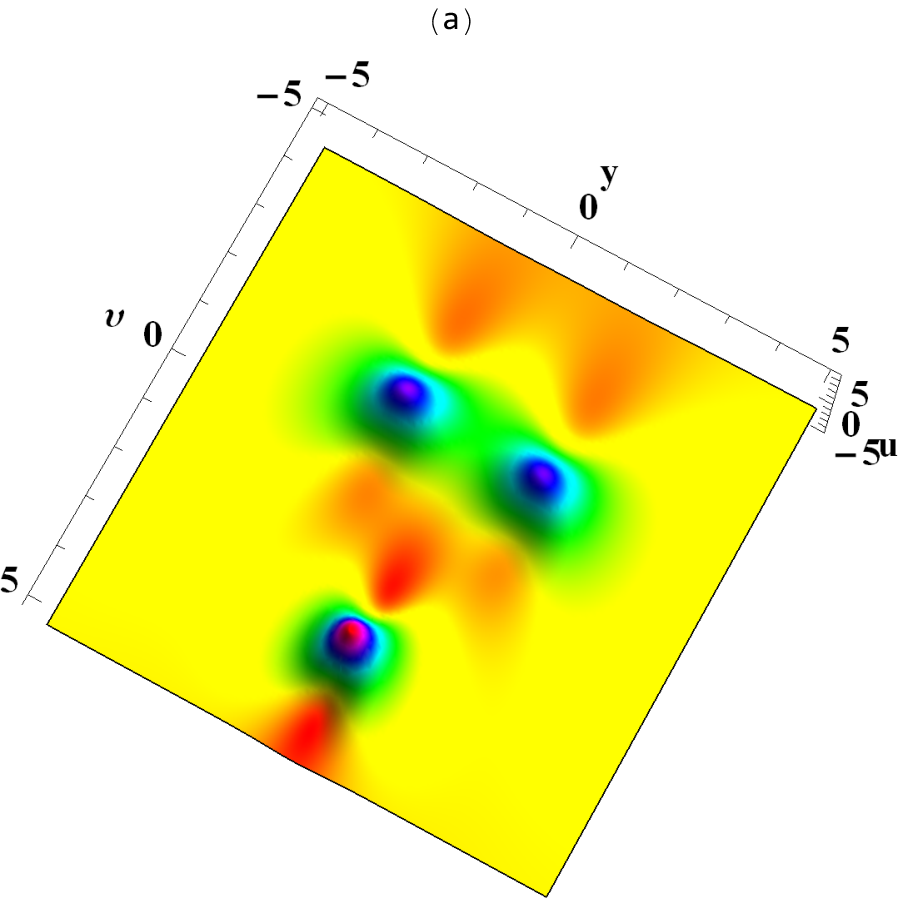}
\includegraphics[scale=0.3,bb=-400 320 10 10]{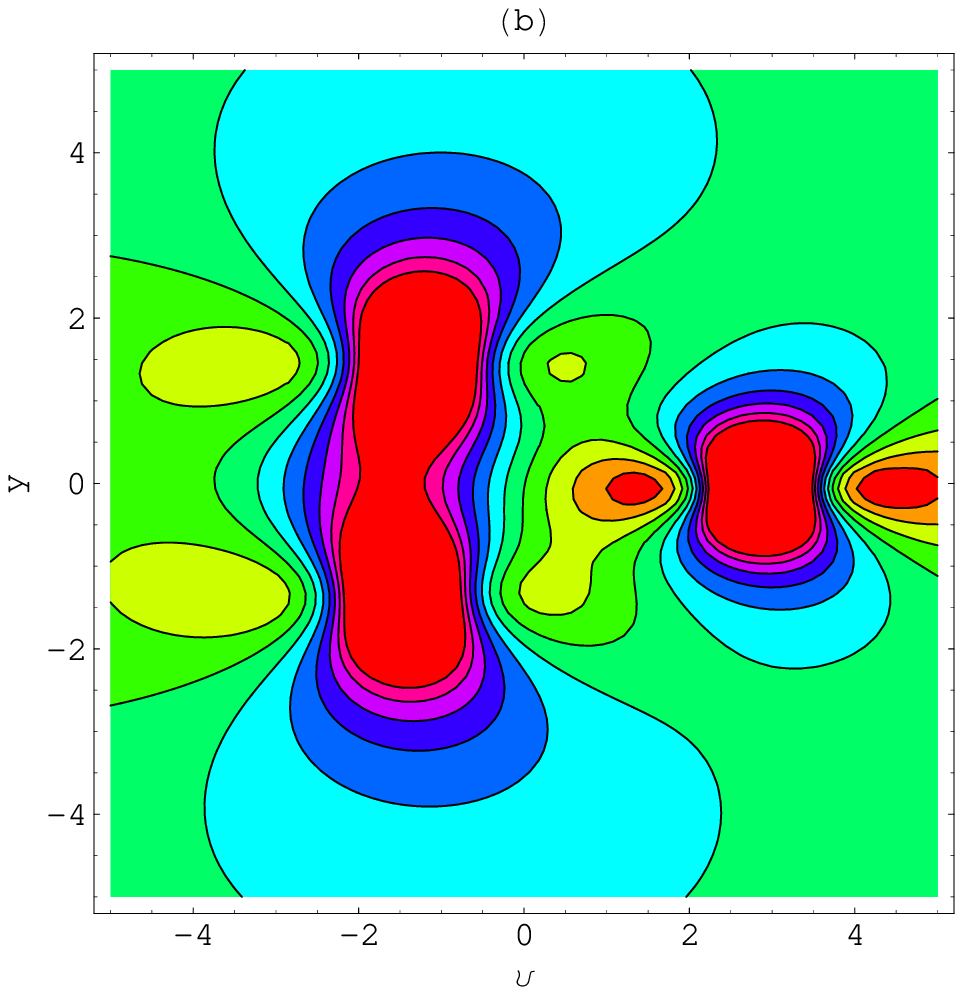}
\includegraphics[scale=0.3,bb=-500 320 10 10]{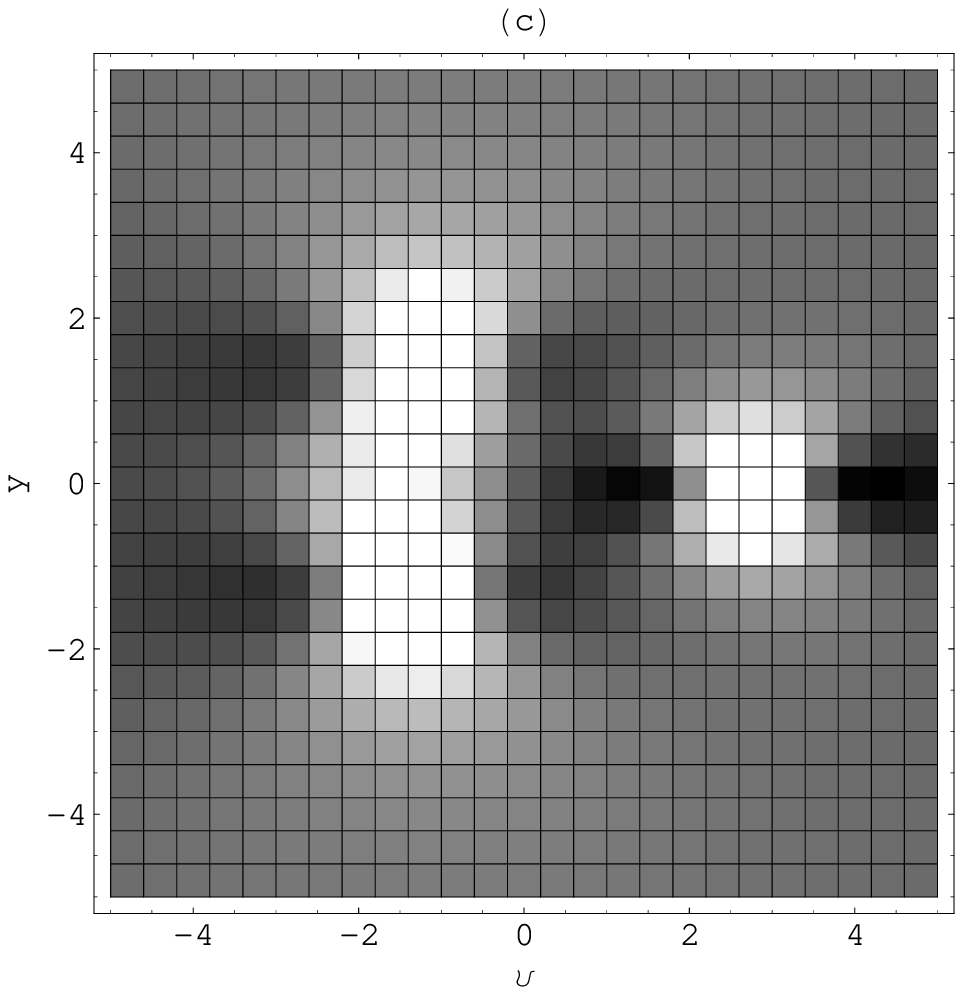}
\vspace{3.2cm}
\begin{tabbing}
\textbf{Fig. 3}. Rogue wave (9) with $\tau_0=0$, $\mu=\nu=10$,
$\vartheta_{19}=\vartheta_{22}=1$, $\alpha=2$,\\ $\Upsilon=-1$, (a)
3D graphic, (b) contour plot, (c) density graphic.
\end{tabbing}

\includegraphics[scale=0.35,bb=20 270 10 10]{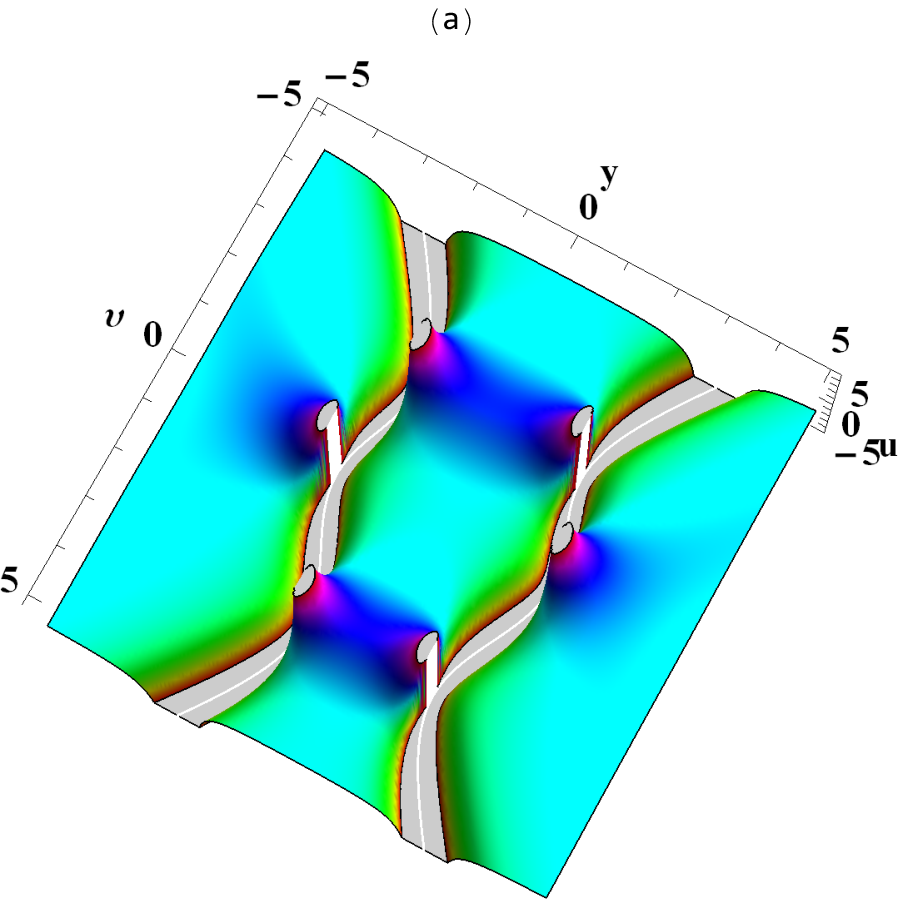}
\includegraphics[scale=0.3,bb=-400 320 10 10]{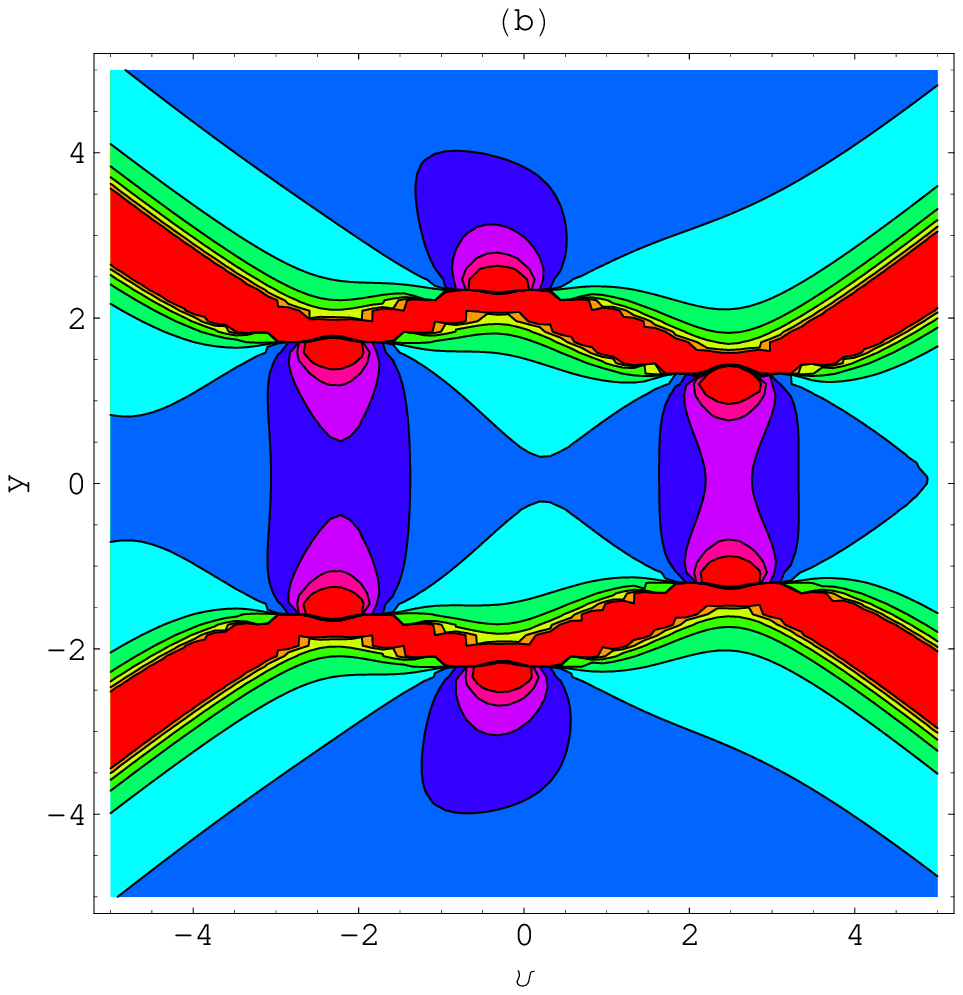}
\includegraphics[scale=0.3,bb=-500 320 10 10]{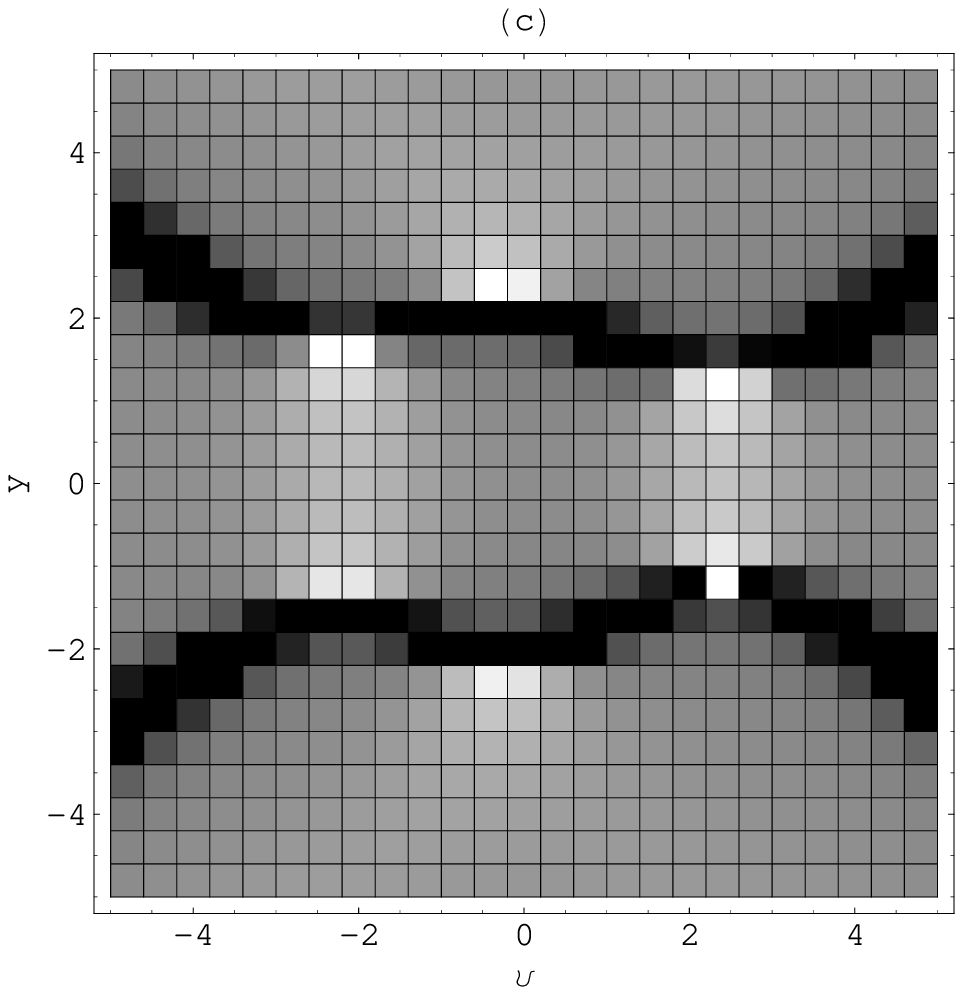}
\vspace{3.2cm}
\begin{tabbing}
\textbf{Fig. 4}. Rogue wave (9) with $\tau_0=0$, $\mu=\nu=10$,
$\vartheta_{19}=\vartheta_{22}=1$, $\alpha=-2$,\\ $\Upsilon=1$, (a)
3D graphic, (b) contour plot, (c) density graphic.
\end{tabbing}
\newpage

\includegraphics[scale=0.35,bb=20 270 10 10]{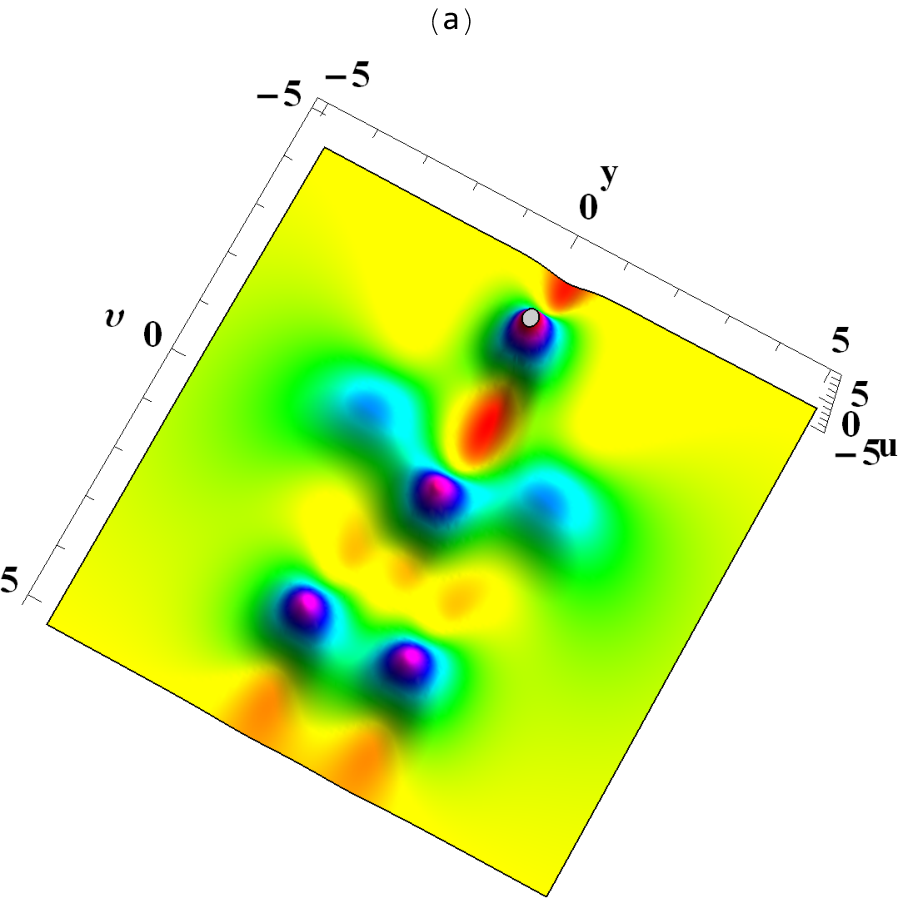}
\includegraphics[scale=0.3,bb=-400 320 10 10]{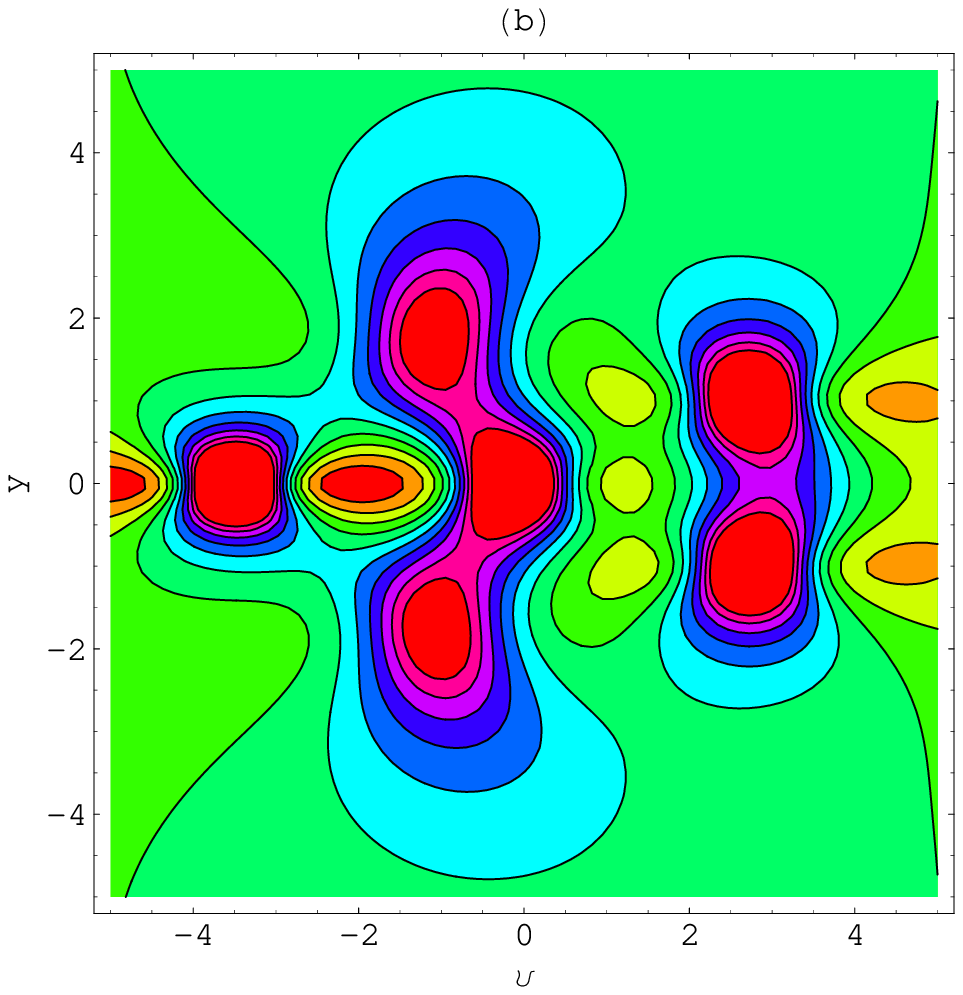}
\includegraphics[scale=0.3,bb=-500 320 10 10]{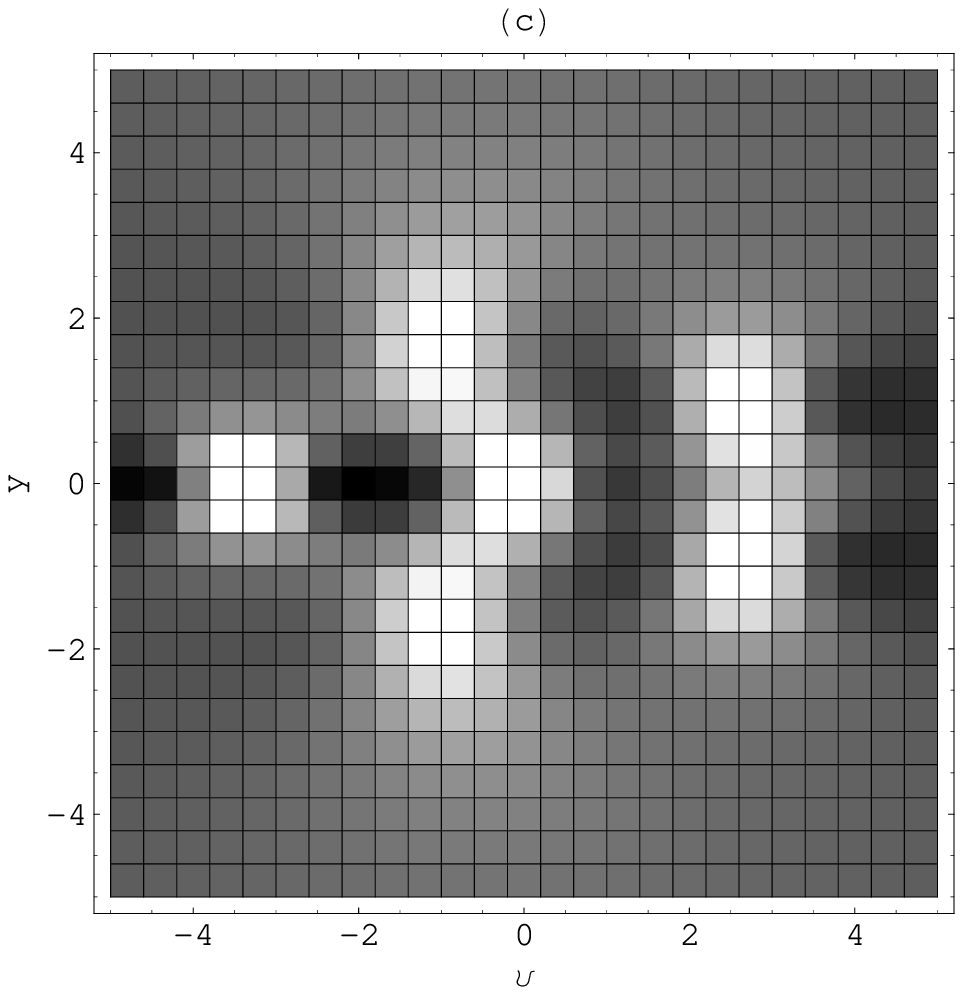}
\vspace{3.2cm}
\begin{tabbing}
\textbf{Fig. 5}. Rogue wave (12) with $\tau_0=0$, $\mu=\nu=1000$,
$\alpha=2$, $\Upsilon=-1$,\\ (a) 3D graphic, (b) contour plot, (c)
density graphic.
\end{tabbing}

\includegraphics[scale=0.35,bb=20 270 10 10]{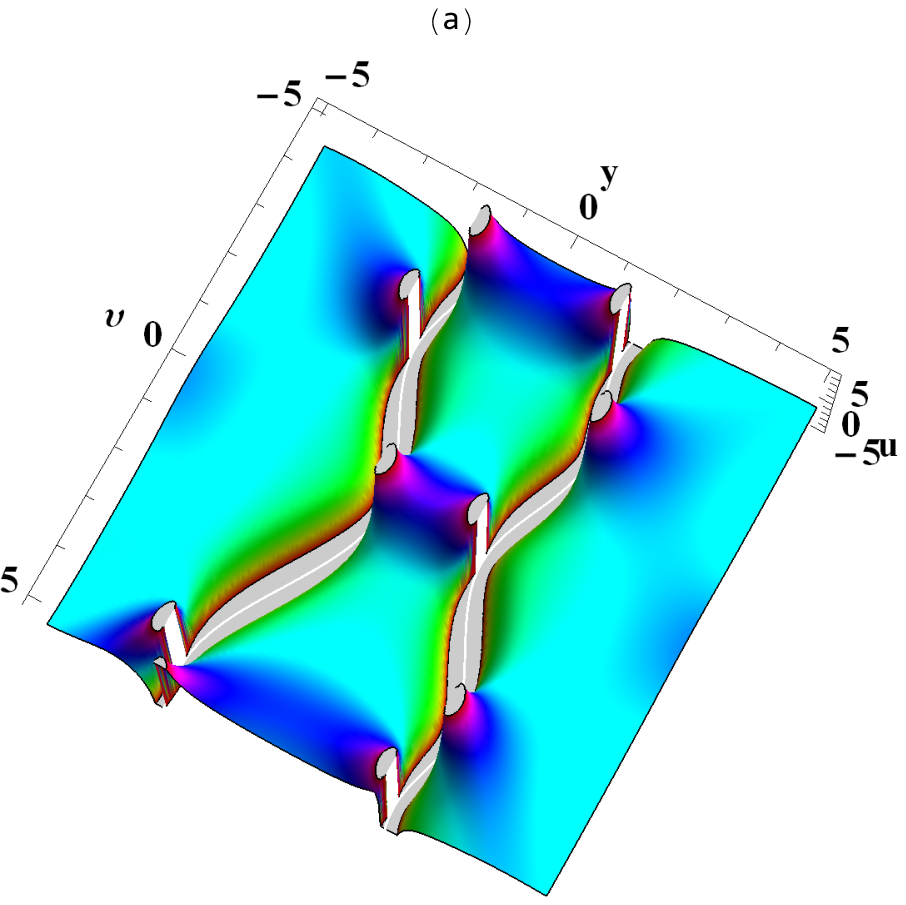}
\includegraphics[scale=0.3,bb=-400 320 10 10]{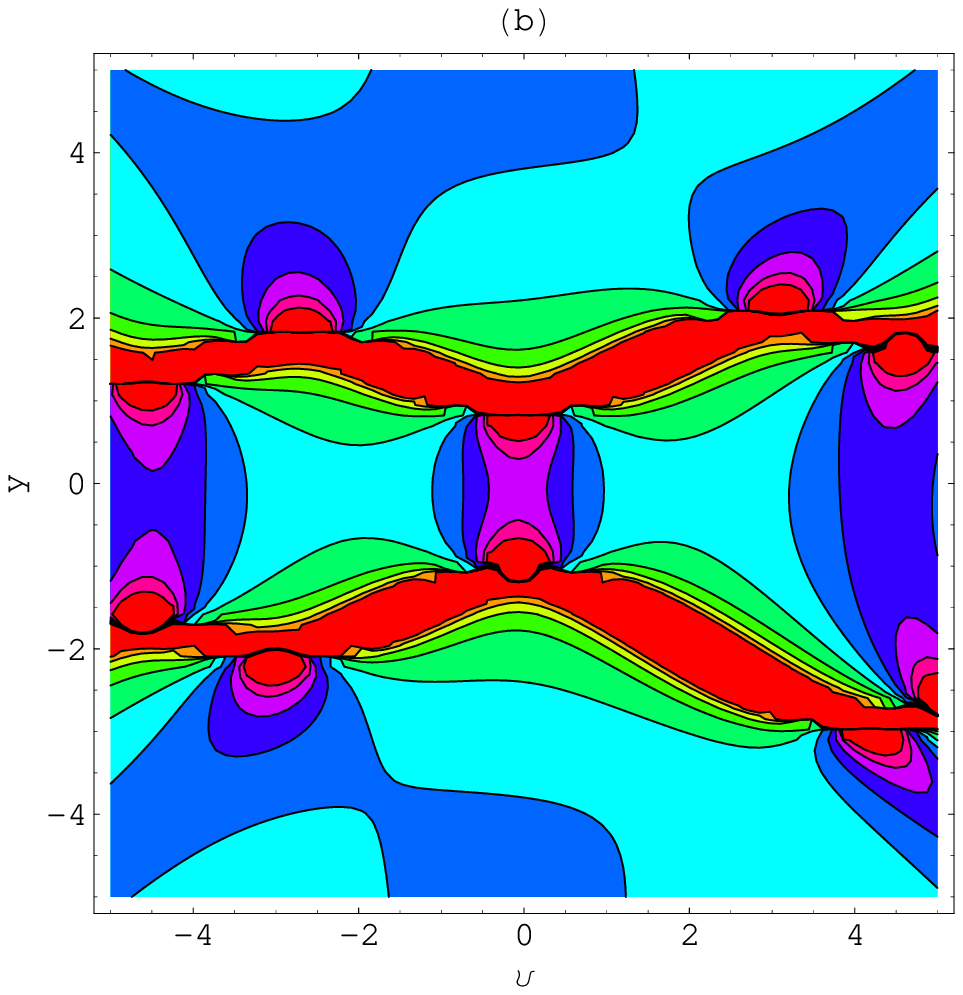}
\includegraphics[scale=0.3,bb=-500 320 10 10]{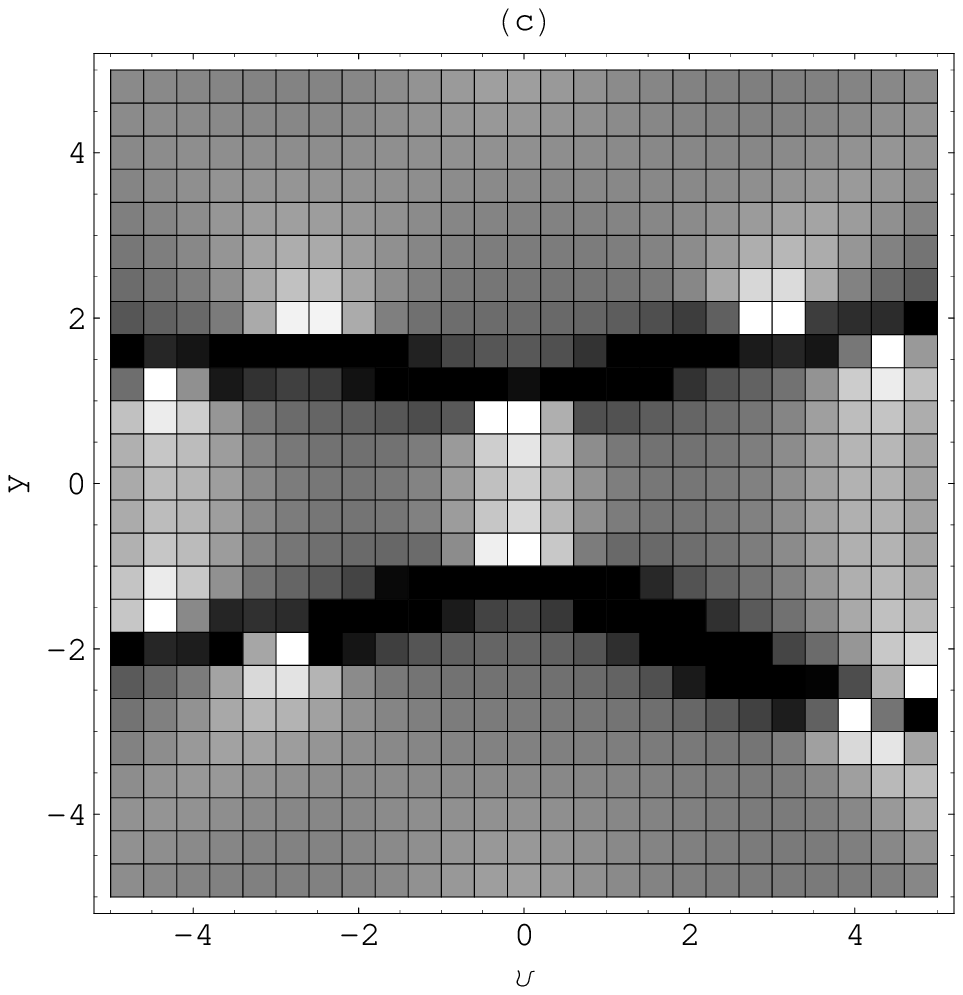}
\vspace{3.2cm}
\begin{tabbing}
\textbf{Fig. 6}. Rogue wave (12) with $\tau_0=0$, $\mu=\nu=1000$,
$\alpha=-2$, $\Upsilon=1$,\\ (a) 3D graphic, (b) contour plot, (c)
density graphic.
\end{tabbing}

\noindent {\bf 4. 6-rogue wave solutions}\\

\quad To present the 6-rogue wave solutions, we choose
{\begin{eqnarray} \xi(\upsilon,y)&=&\upsilon ^{12}+y^8 \vartheta
_{48}+y^6 \vartheta
   _{47}+y^4 \vartheta _{46}+\upsilon ^{10}
   \left(y^2 \vartheta _{26}+\vartheta _{25}\right)\nonumber\\&+&y^2 \vartheta _{45}+\upsilon ^8 \left(y^4
   \vartheta _{29}+y^2 \vartheta _{28}+\vartheta _{27}\right)+2 \mu  \upsilon  [\upsilon ^6+y^6
   \vartheta _{64}+y^4 \vartheta _{63}\nonumber\\&+&\upsilon ^4 \left(y^2 \vartheta _{69}+\vartheta
   _{68}\right)+y^2 \vartheta _{62}+\upsilon ^2 \left(y^4 \vartheta _{67}+y^2 \vartheta
   _{66}+\vartheta _{65}\right)+\vartheta _{61}]\nonumber\\&+&2 \nu  y [y^6+y^4 \left(\upsilon ^2
   \vartheta _{57}+\vartheta _{56}\right)+y^2 \left(\upsilon ^4 \vartheta _{55}+\upsilon ^2
   \vartheta _{54}+\vartheta _{53}\right)+\upsilon ^6 \vartheta _{60}\nonumber\\&+&\upsilon ^4 \vartheta
   _{59}+\upsilon ^2 \vartheta _{58}+\vartheta _{52}]+\upsilon ^6 \left(y^6 \vartheta
   _{33}+y^4 \vartheta _{32}+y^2 \vartheta _{31}+\vartheta _{30}\right)\nonumber\\&+&\upsilon ^4 \left(y^8
   \vartheta _{38}+y^6 \vartheta _{37}+y^4 \vartheta _{36}+y^2 \vartheta _{35}+\vartheta
   _{34}\right)+\upsilon ^2 (y^{10} \vartheta _{44}+y^8 \vartheta _{43}\nonumber\\&+&y^6 \vartheta _{42}+y^4
   \vartheta _{41}+y^2 \vartheta _{40}+\vartheta _{39})+\vartheta _{51}+y^{12} \vartheta _{50}+y^{10}
\vartheta _{49}\nonumber\\&+&\left(\mu ^2+\nu ^2\right) [-\frac{3
\Upsilon }{\alpha +6 \tau _0
   \Upsilon +1}+\upsilon ^2+y^2 \left(\alpha +6 \tau _0 \Upsilon
   +1\right)],
\end{eqnarray}
}where $\vartheta _i (i=25,\cdots, 69)$  is unknown real constant.
Substituting Eq. (10) into Eq. (3) and equating the coefficients of
all powers $\upsilon$ and $y$ to zero, we get
\begin{eqnarray}
\vartheta_{26}&=&6 \left(\alpha +6 \tau _0 \Upsilon +1\right),
\vartheta_{29}=15 \left(\alpha +6 \tau _0 \Upsilon +1\right){}^2,
\vartheta_{28}=-690 \Upsilon,\nonumber\\ \vartheta_{33}&=&20
\left(\alpha +6 \tau _0 \Upsilon +1\right){}^3,
\vartheta_{50}=\left(\alpha +6 \tau _0 \Upsilon +1\right){}^6,
\vartheta_{44}=6 \left(\alpha +6 \tau _0 \Upsilon +1\right){}^5,\nonumber\\
\vartheta_{38}&=&15 \left(\alpha +6 \tau _0 \Upsilon +1\right){}^4,
\vartheta_{32}=-1540 \Upsilon \left(\alpha +6 \tau _0 \Upsilon
+1\right), \vartheta_{36}=37450 \Upsilon ^2,\nonumber\\
\vartheta_{31}&=&\frac{18620 \Upsilon ^2}{\alpha +6 \tau _0 \Upsilon
+1}, \vartheta_{49}=-58 \Upsilon \left(\alpha +6 \tau _0 \Upsilon
+1\right){}^4, \vartheta_{58}=-\frac{665 \Upsilon
^2}{\left(\alpha +6 \tau _0 \Upsilon +1\right){}^5},\nonumber\\
\vartheta_{37}&=&-1460 \Upsilon \left(\alpha +6 \tau _0 \Upsilon
+1\right){}^2, \vartheta_{55}=-\frac{5}{\left(\alpha +6 \tau _0
\Upsilon +1\right){}^2},\nonumber\\ \vartheta_{43}&=&-570 \Upsilon
\left(\alpha +6 \tau _0 \Upsilon +1\right){}^3, \vartheta_{48}=4335
\Upsilon ^2 \left(\alpha +6 \tau _0 \Upsilon
+1\right){}^2,\end{eqnarray}\begin{eqnarray}
\vartheta_{35}&=&-\frac{220500 \Upsilon ^3}{\left(\alpha +6 \tau _0
\Upsilon +1\right){}^2}, \vartheta_{42}=35420 \Upsilon ^2
\left(\alpha +6 \tau _0 \Upsilon +1\right),\nonumber\\
\vartheta_{41}&=&\frac{14700 \Upsilon ^3}{\alpha +6 \tau _0 \Upsilon
+1}, \vartheta_{54}=\frac{190 \Upsilon }{\left(\alpha +6 \tau _0
\Upsilon +1\right){}^3}, \vartheta_{57}=-\frac{9}{\alpha +6 \tau _0
\Upsilon +1},\nonumber\\ \vartheta_{40}&=&\frac{565950 \Upsilon
^4}{\left(\alpha +6 \tau _0 \Upsilon +1\right){}^3},
\vartheta_{64}=5 \left(\alpha +6 \tau _0 \Upsilon +1\right){}^3,
\vartheta_{47}=-\frac{798980 \Upsilon ^3}{3},\nonumber\\
\vartheta_{67}&=&-5 \left(\alpha +6 \tau _0 \Upsilon +1\right){}^2,
\vartheta_{63}=-45 \Upsilon  \left(\alpha +6 \tau _0 \Upsilon
+1\right),\nonumber\\ \vartheta_{25}&=&-\frac{98 \Upsilon }{\alpha
+6 \tau _0 \Upsilon +1}, \vartheta_{69}=-9 \left(\alpha +6 \tau _0
\Upsilon +1\right), \vartheta_{68}=-\frac{13 \Upsilon }{\alpha +6
\tau _0 \Upsilon +1},\nonumber\\ \vartheta_{62}&=&\frac{535 \Upsilon
^2}{\alpha +6 \tau _0 \Upsilon +1}, \vartheta_{65}=-\frac{245
\Upsilon ^2}{\left(\alpha +6 \tau _0 \Upsilon +1\right){}^2},
\vartheta_{46}=\frac{16391725 \Upsilon ^4}{3 \left(\alpha +6 \tau _0
\Upsilon +1\right){}^2},\nonumber\\ \vartheta_{56}&=&\frac{7
\Upsilon }{\left(\alpha +6 \tau _0 \Upsilon +1\right){}^2},
\vartheta_{53}=-\frac{245 \Upsilon ^2}{\left(\alpha +6 \tau _0
\Upsilon +1\right){}^4}, \vartheta_{60}=\frac{5}{\left(\alpha +6
\tau _0 \Upsilon +1\right){}^3},\nonumber\\
\vartheta_{27}&=&\frac{735 \Upsilon ^2}{\left(\alpha +6 \tau _0
\Upsilon +1\right){}^2}, \vartheta_{34}=-\frac{5187875 \Upsilon
^4}{3 \left(\alpha +6 \tau _0 \Upsilon +1\right){}^4},\nonumber\\
\vartheta_{59}&=&-\frac{105 \Upsilon }{\left(\alpha +6 \tau _0
\Upsilon +1\right){}^4},
\vartheta_{61}=-\frac{12005 \Upsilon ^3}{3 \left(\alpha +6 \tau _0 \Upsilon +1\right){}^3}, \vartheta_{66}=230 \Upsilon,\nonumber\\
\vartheta_{52}&=&-\frac{18865 \Upsilon ^3}{3 \left(\alpha +6 \tau _0
\Upsilon +1\right){}^6}, \vartheta_{30}=-\frac{75460 \Upsilon ^3}{3
\left(\alpha +6 \tau _0 \Upsilon +1\right){}^3},\nonumber\\
\vartheta_{51}&=&\frac{3 \nu ^2 \Upsilon [\left(\alpha +6 \tau _0
\Upsilon +1\right){}^7-1]}{\left(\alpha
   +6 \tau _0 \Upsilon +1\right){}^8}+\frac{878826025 \Upsilon ^6}{9 \left(\alpha +6 \tau _0
   \Upsilon +1\right){}^6},\nonumber\\
\vartheta_{39}&=&\nu ^2 [\frac{1}{\left(\alpha +6 \tau _0 \Upsilon
+1\right){}^7}-1]-\frac{159786550
   \Upsilon ^5}{3 \left(\alpha +6 \tau _0 \Upsilon +1\right){}^5},
   \nonumber\\
    \vartheta_{45}&=&\nu ^2 [\frac{1}{\left(\alpha +6 \tau _0 \Upsilon +1\right){}^6}-\alpha
   -1]-\frac{300896750 \Upsilon ^5}{3 \left(\alpha +6 \tau _0 \Upsilon +1\right){}^4}-6 \nu ^2
   \tau _0 \Upsilon.
\end{eqnarray}Substituting Eq. (10) and Eq. (11) into Eq. (2), the 6-rogue wave
solutions for Eq. (1) can be obtained as {\begin{eqnarray}
u=-\frac{2 \xi_\upsilon^2}{\xi ^2}+\frac{2
\xi_{\upsilon\upsilon}}{\xi
   }+\tau _0,
\end{eqnarray}}where $\xi$ satisfies Eq. (10) and Eq. (11). Dynamics features of
6-rogue wave solutions are shown in Fig. 5 ($\Upsilon<0$) and Fig.6
($\Upsilon>0$). It can be found that there are six 1-rogue wave
solutions.

\noindent {\bf 5. Conclusion}\\

\quad In the paper,  the gCHKP equation is investigated.  Multiple
rogue wave solutions are discussed by using the symbolic computation
approach. As an result, the 1-rogue wave solutions, 3-rogue wave
solutions and 6-rogue wave solutions are presented, respectively.
Meantime, their dynamics features are displayed in Figs. 1-6. From
the above calculation process,
 the symbolic computation approach is more useful and direct,
 but the calculation amount is very large, which needs the help of Mathematical software.


\begin{thebibliography}{s20}
\bibitem{s1} Clarkson, P.A.,  Dowie, E.: Rational solutions of the Boussinesq
equation and applications to rogue waves. Trans. Math. Appl., 1(1),
 tnx003 (2017).
\bibitem{s1}  Dysthe, K.,  Krogstad, H.E.,   Muller, P.: Oceanic rogue waves. Annu. Rev. Fluid Mech., 40,  287-310 (2008).
\bibitem{s1}  Dudley, J.M.,  Dias, F.,  Erkintalo, M.,   Genty, G.: Instabilities, breathers and rogue waves in optics. Nat. Photonics,
8, 755-764 (2014).
\bibitem{s2} Ma, Y.L., Li, B.Q.: Interactions between soliton and rogue wave for a (2+1)-dimensional
generalized breaking soliton system: Hidden rogue wave and hidden
soliton. Comput. Math. Appl., 78(3), 827-839 (2019)
\bibitem{s3} Lan, Z.Z.: Rogue wave solutions for a coupled nonlinear Schr?dinger equation in the birefringent optical
fiber. Appl. Math. Lett., 98, 128-134 (2019).
\bibitem{s4} Geng, X.G., Shen, J., Xue, B.: Dynamical behaviour of rogue wave solutions in a multi-component AB
system. Wave Motion, 89, 1-13 (2019).
\bibitem{s1} Mao, J.J., Tian, S.F., Zhang, T.T.: Rogue waves, homoclinic breather waves and soliton waves for a
 (3+1)-dimensional non-integrable KdV-type equation. Int. J. Numer. Method. H.,
 29(2),763-772.
\bibitem{s5} Jia, R.R., Guo, R.: Breather and rogue wave solutions for the (2+1)-dimensional nonlinear Schr\"{o}dinger-Maxwell-Bloch
equation. Appl. Math. Lett., 93, 117-123 (2019).
\bibitem{s6} Gao, L.N., Zhao, X.Y., Zi, Y.Y., Yu, J., L\"{u}, X.: Resonant behavior of multiple wave solutions to
a Hirota bilinear equation. Comput. Math. Appl., 72, 1225-1229
(2016).
\bibitem{s7} Gao, L.N., Zi, Y.Y.,  Yin, Y.H.,  Ma, W.X., L\"{u}, X.: B\"{a}cklund transformation,
 multiple wave solutions and lump solutions to a (3 + 1)-dimensional nonlinear evolution equation. Nonlinear Dyn., 89, 2233-2240  (2017).
\bibitem{s11}  Wazwaz, A.M.: The integrable time-dependent sine-Gordon equation with multiple optical kink solutions.
Optik, 182, 605-610  (2019).
\bibitem{s12}   Yin, Y.H.,  Ma, W.X.,  Liu, J.G.,  L\"{u}, X.: Diversity of exact solutions to a (3+1)-dimensional nonlinear evolution equation and its reduction.
 Comput. Math. Appl., 76,  1275-1283 (2018).
\bibitem{s13}  Wazwaz, A.M., Kaur, L.: New integrable Boussinesq equations of distinct
dimensions with diverse variety of soliton solutions. Nonlinear
Dyn., 97(1), 83-94 (2019).
\bibitem{s14} Xu, G.Q.,  Wazwaz, A.M.: Characteristics of integrability, bidirectional solitons and
localized solutions for a (3 + 1)-dimensional generalized breaking
soliton equation. Nonlinear Dyn.,  96, 1989-2000 (2019).
\bibitem{s15}   Liu, J.G.,  Zhu, W.H.: Breather wave solutions for the generalized shallow water wave
 equation with variable coefficients in the atmosphere, rivers, lakes and oceans. Comput. Math. Appl., 78(3),  848-856 (2019).
\bibitem{s16}   He, J.S.,  Xu, S.W.,  Cheng, Y.:  The rational solutions of the mixed nonlinear Schr\"{o}dinger equation. AIP Adv., 5(1), 017105 (2014).
\bibitem{s19}  Osman, M.S., Inc, M., Liu, J.G., Hosseini, K.,  Yusu,
A.: Different wave structures and stability analysis for the
generalized (2+1)-dimensional Camassa-Holm-Kadomtsev-Petviashvili
equation. Phys. Scripta, 95(3), 035229 (2020).
\bibitem{s20} Biswas, A.: 1-Soliton solution of the generalized Camassa-Holm Kadomtsev-Petviashvili
equation. Commun. Nonlinear Sci. Numer. Simulat., 14, 2524-2527
(2009).
\bibitem{s21} Ebadi, G., Yousefzadeh Fard, N., Triki, H., Biswas, A.: Exact solutions of the (2+1)-
dimensional Camassa-Holm Kadomtsev-Petviashvili equation. Nonlinear
Anal. Model. Control, 17, 280-296 (2012).
\bibitem{s18} Zhen-Li, W., Xi-Qiang, L.: Symmetry reductions and exact solutions of the (2+1)-
dimensional Camassa-Holm Kadomtsev-Petviashvili equation. Pramana,
85, 3-16 (2015).
\bibitem{s23} Qin, C. Y., Tian, S. F., Wang, X. B., Zhang, T. T.: On breather waves, rogue waves
and solitary waves to a generalized (2+1)-dimensional
Camassa-Holm-Kadomtsev- Petviashvili equation. Commun. Nonlinear
Sci. Numer. Simulat., 62, 378-385 (2018).
\bibitem{s18} Zha,Q.L.: A symbolic computation approach to constructing rogue waves with a controllable center in the nonlinear
systems. Comput. Math. Appl., 75(9),  3331-3342 (2018).
\bibitem{s18} Liu, W.H., Zhang, Y.F.: Multiple rogue wave solutions for a (3+1)-dimensional Hirota
bilinear equation. Appl. Math. Lett., 98,  184-190 (2019).
\bibitem{s26}  Gaillard, P.: Rational solutions to the KPI equation and multi rogue waves. Ann. Phys., 367,  1-5 (2016).
\bibitem{s29} Zhao, Z.L., He, L.C., Gao, Y.B.: Rogue Wave and Multiple Lump Solutions of
the (2+1)-Dimensional Benjamin-Ono Equation in Fluid Mechanics.
Complexity,, 2019, 8249635 (2019).
\bibitem{s29} Zhao, Z.L., He, L.C.: Multiple lump solutions of the (3+1)-dimensional potential
Yu-Toda-Sasa-Fukuyama equation. Appl. Math. Lett., 95, 114-121
(2019).

\end{thebibliography}
\end{document}